\documentclass[aps,pra,reprint,superscriptaddress,showpacs,floatfix]{revtex4-1}
\usepackage{float}
\usepackage{amsfonts}
\usepackage{amsmath}
\usepackage{amssymb}
\usepackage{graphicx}
\usepackage{listings}
\usepackage{bm}
\usepackage{bbold}
\usepackage{braket}
\usepackage{todonotes}
\usepackage{hyperref}
\usepackage{array}
\usepackage{multirow}
\usepackage{url}
\usepackage[utf8]{inputenc}

\newcolumntype{C}[1]{>{\centering\let\newline\\\arraybackslash\hspace{0pt}}m{#1}}
\newcommand{\Hb}{\overline{\mathrm{H}}}

\begin{document}

\title{Quantum Reflection of Antihydrogen from Nanoporous Media}

\author{G. Dufour}
\affiliation{Laboratoire Kastler-Brossel, CNRS, ENS, UPMC, Campus
Jussieu, F-75252 Paris, France}
\author{R. Gu{\'e}rout}
\affiliation{Laboratoire Kastler-Brossel, CNRS, ENS, UPMC, Campus
Jussieu, F-75252 Paris, France}
\author{A. Lambrecht}
\affiliation{Laboratoire Kastler-Brossel, CNRS, ENS, UPMC, Campus
Jussieu, F-75252 Paris, France}
\author{V.V. Nesvizhevsky}
\affiliation{Institut Laue-Langevin (ILL), 6 rue Jules Horowitz,
 F-38042, Grenoble, France}
\author{S. Reynaud}
\affiliation{Laboratoire Kastler-Brossel, CNRS, ENS, UPMC, Campus
Jussieu, F-75252 Paris, France}
\author{A.Yu. Voronin}
\affiliation{ P.N. Lebedev Physical Institute, 53 Leninsky prospect,
Ru-117924 Moscow, Russia}

\begin{abstract}
We study quantum reflection of antihydrogen atoms from nanoporous
media due to the Casimir-Polder (CP) potential.
Using a simple effective medium model, we show a dramatic increase
of the probability of quantum reflection of antihydrogen atoms if the
porosity of the medium increases. We discuss the limiting case of
reflections at small energies, which have interesting applications
for trapping and guiding antihydrogen using material walls.
\end{abstract}


\maketitle

\section{Introduction}

In a previous paper \cite{dufour2012} we have shown that
antihydrogen atoms have a significant probability of being reflected
when they fall on a material surface. This quantum reflection occurs
as a consequence of the rapid variation of the Casimir-Polder (CP)
interaction on the scale of the atomic de Broglie wavelength \cite{berry1972,yu1993,shimizu2001,druzhinina2003,oberst2005,cronin2009,zhao2010,pasquini2004,pasquini2006,friedrich2002,judd2011}.

We did emphasize in \cite{dufour2012} that quantum reflection
increased for weaker CP potentials. The explanation is that quantum
reflection takes place closer to the surface for weaker potentials,
and that the potential is steeper there, leading to a greater
probability of non-adiabatic transition. As the CP interaction is
due to the coupling of atoms with surface through electromagnetic
vacuum fluctuations, a weaker coupling of the surface to the
electromagnetic field leads to increased quantum reflection. This
paradoxical result was discussed in \cite{dufour2012} 
%gabriel replaced ``for'' by ``in relation to''
in relation to the effect
of the dielectric index, as well as that of thickness for matter
slabs. Here, we wish to extend this work by focusing on the
enhancement of quantum reflection from nanoporous materials.

Increasing the reflection probability of antihydrogen would open the
possibility for trapping and guiding antimatter with material walls, in particular
anti-atoms can be trapped in gravitational quantum states held
by gravity from above and quantum reflection from below
\cite{voronin2011,voronin2012}. Such states are also of
interest for the GBAR experiment which is designed to measure the
gravitational properties of antimatter \cite{gbar}. In the current
design, the acceleration of gravity $\overline{g}$ for antihydrogen
atoms is measured by timing their fall from the release from a trap
to the annihilation at contact with a matter plate. In the future,
the acceleration of gravity $\overline{g}$ could be measured with a
better precision through a spectroscopic study of gravitational
quantum states.

In the present paper, we will consider several examples of
nanoporous media~: silica aerogels, porous silicon and powders of
diamond nanoparticles. These materials incorporate a significant
fraction of gas or vacuum, which reduces their effective
permittivity. We expect a significant increase in quantum reflection
probability when the average density of the medium is significantly
smaller than the bulk density of the constituent material. As
inhomogeneities in such materials have a nanometric scale, we may
describe the material by an effective medium theory and still expect
approximately correct results for processes involving larger scales.

Silica aerogels are generally produced by supercritically drying a
silica gel, a process which removes the liquid component while
leaving the silica matrix undamaged. The result is a disordered
array of pores in a solid framework whose properties such as
porosity (volume fraction of gas or vacuum) and pore size are
tunable. Porosities as high as 98\% are obtained for aerogels, with
pore sizes typically of the order of 20-100 nm \cite{sinko2010}.
Porous silicon can be obtained by anodization or 
%gabriel removed ``stain'' (not sure what it is, prefer to remain general)
etching of a
silicon wafer. Porosities up to 95\% are obtained with pore sizes
typically ranging from 2 to 50 nm \cite{granitzer2010,bisi2000}.

%\todo[inline]{find better reference for porous silicon? }

Diamond nanoparticles formed by explosive shock have been observed
for more than forty years \cite{decarli1961}. Intensive studies of
their fabrication and applications have been performed worldwide
\cite{dolmatov2007}. These particles, with typical sizes of a few
nanometers, consist of a diamond nucleus (with diamond density)
within a thin onion-like shell with complex chemical composition and
lower density \cite{aleksenskii1999}. Powders of diamond
nanoparticles have been intensively studied and used in neutron
investigations in relation to their unique property of extremely
efficient reflection of slow neutrons
% \cite{nesvizhevsky2002,nesvizhevsky2008,lychagin2009,nesvizhevsky2010,cubitt2010,krylov2011,nesvizhevsky2012quantum}
\cite{nesvizhevsky2008,nesvizhevsky2010}.
The analogy between the interaction of slow antihydrogen atoms and
neutrons with nanodiamond powders (via the CP and Fermi interactions
respectively) is underlined by the fact that their masses, typical
velocities and wavelengths are equal. The typical density in powder
varies between 5\% and 15\% of the diamond density, depending on the
conditioning of the powder; it can be increased by shaking or
pressing the powder, and decreased by blowing air through it. The
density close to surface of a nanoparticle powder sample is even
lower than that the bulk density of the powder (see for instance
\cite{nesvizhevsky2012study}).

Quantum reflection experiments have been performed by Pasquini et al
with condensates of sodium atoms on a nano-structured silicon
surface and a silica aerogel \cite{pasquini2006}. A reflection
probability of $\sim$60\% was observed on the nano-structured
silicon surface, compared with $\sim$ 15\% on bulk silicon. However,
no reflection was observed on the aerogel.
%gabriel added a full stop here
 The authors of
\cite{pasquini2006} impute this observation to uncontrolled surface
charges. This is a serious issue to be solved in order to use the
spectacular enhancement of reflection predicted in the present
paper.

The dielectric properties of the material over a broad range of
frequencies are needed to compute the CP force. Though a porous
medium is by nature inhomogeneous, we will use an effective medium
approximation and describe the composite material as an homogeneous
medium with an effective permittivity $\epsilon$. We will also make
the simplifying assumption of a plane interface, whereas a porous
medium is likely to have a rough surface. Such approximations are
valid if the wavelengths involved are larger than the scales of
inhomogeneity or roughness. Qualitatively, one expects our
simplified description to lead to approximately correct results if
the atom does not approach the medium closer than these scales. It
will therefore be sufficient for a first exploration of the
enhancement of quantum reflection by the use of low-density
materials. A more complete description will require further work, as
it would have to study the CP effect in presence of non specular
scattering of electromagnetic field \cite{contreras2010}
as well as the resulting non specular quantum reflection of
antihydrogen.

%\todo[inline]{reference for non specular quantum reflection? }

\section{Effective medium model}

There are several models available for obtaining the effective
permittivity $\epsilon$ as a function of the permittivities of the
constituent materials. For a host material containing
non-overlapping spherical inclusions of another material,
application of the Clausius-Mossoti relation yields the
Maxwell-Garnett formula\cite{maxwellgarnett1904}:
\begin{equation}
\frac{\epsilon-\epsilon_h}{\epsilon+2\epsilon_h}=
\phi_i \frac{\epsilon_i-\epsilon_h}{\epsilon_i+2\epsilon_h}
\end{equation}
where $\epsilon_h$ and $\epsilon_i$ are the permittivity of the host
and inclusions respectively and $\phi_i$ is the volume fraction of
inclusions.

Another possibility is the Bruggeman model \cite{bruggeman1935},
which requires that the average polarization of spherical inclusions
embedded in the effective medium vanish. For a mixture of materials
with permittivities $\epsilon_1$ and $\epsilon_2$ and volume
fractions $\phi_1$ and $\phi_2$ ($\phi_1+\phi_2=1$), the effective
permittivity $\epsilon$ is given by:
\begin{equation}
\phi_1 \frac{\epsilon_1-\epsilon}{\epsilon_1+2\epsilon}
+\phi_2 \frac{\epsilon_2-\epsilon}{\epsilon_2+2\epsilon}=0
\end{equation}

Landau and Lifshitz have proposed another effective model based on
the idea that the cubic root of the permittivity is approximately
additive \cite{landau-continuousmedia}:
\begin{equation}
\epsilon^{1/3}=\phi_1 \ \epsilon_1^{1/3}+\phi_2 \ \epsilon_2^{1/3}
\end{equation}

\begin{figure}[ht]
\centering
\includegraphics[width=8cm]{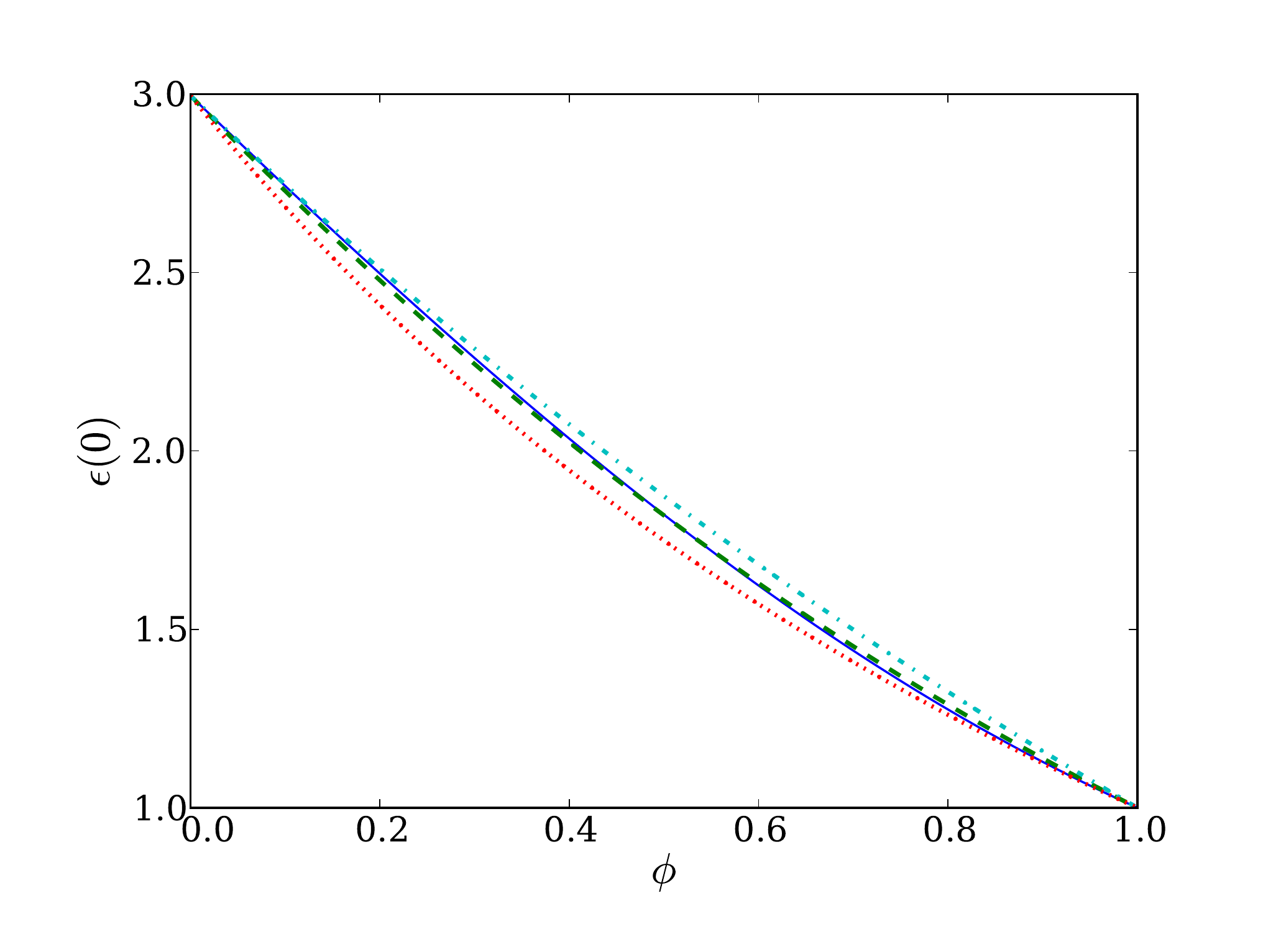}
\caption{\label{effective}(Color online) Comparison of effective
medium models for the effective static permittivity of a silica
aerogel as a function of porosity: Bruggeman (blue full line),
Landau-Lifschitz (green dashed line), Maxwell-Garnett with air
inclusions (cyan dot-dashed line), Maxwell-Garnett with silica
inclusions (red dotted line).}
\end{figure}

In the Casimir-related literature, aerogels have been described by
Maxwell-Garnett models with air inclusions in a solid matrix
\cite{esquivel2007} or vice-versa \cite{bimonte2011}. Note that this
Maxwell-Garnett model does not treat the host and the inclusions in
a symmetrical manner. Here we prefer to use a symmetrical model,
like the Landau-Lifshitz and Bruggeman models, since we want to vary
the porosity over a wide range of values. As shown in figure
\ref{effective}, the Landau-Lifshitz and Bruggeman models give very
similar results and we chose the latter for its simpler
interpretation. In the rest of this paper, we denote $\phi$ the
porosity, i.e. the volume fraction of gas or vacuum, and $1-\phi$
the solid fraction, that is also the density reduction.

The optical response properties used for antihydrogen, silica and silicon are the same as in \cite{dufour2012}.
Diamond is described by a simple Sellmeier model \cite{ghosh1998}:
\begin{equation}
\label{diamond} \epsilon(i\xi)=B_0+\frac{B_1}{1+(\xi/\omega_1)^2}+\frac{B_2}{1+(\xi/\omega_2)^2}~,
\end{equation}
with the parameters $B_{0,1,2}$= 2.30982863, 3.35656148, 3.25669602 and
$\omega_{1,2}=$ 14.3235, 0.0376730 $\times$ 10$^{15}$rad.s$^{-1}$.

\section{Casimir-Polder potential}

%gabriel replaced Casimir-Polder by its abbreviation here and in the figure captions below

\begin{figure}[ht]
\centering
\includegraphics[width=8cm]{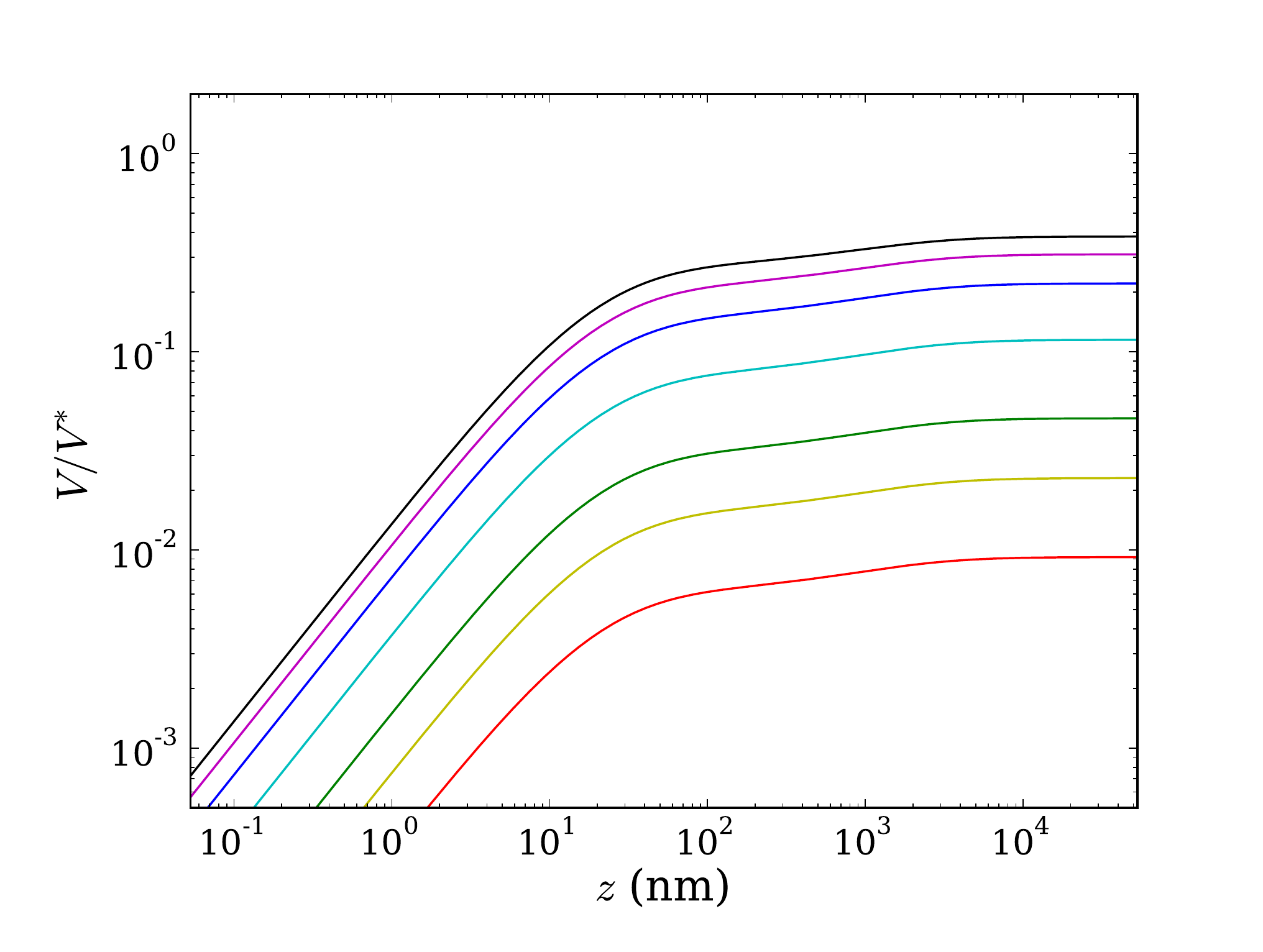}
\caption{\label{vcp-aero}(Color online) CP potential $V$
for antihydrogen on silica aerogels, divided by the long-range
potential $V^*$ for a perfect mirror. From top to bottom the
porosity is 0\% (bulk silica, black), 25\% (magenta), 50\% (blue),
75\% (cyan), 90\% (green), 95\% (yellow), 98\% (red).}
\end{figure}

\begin{figure}[ht]
\centering
\includegraphics[width=8cm]{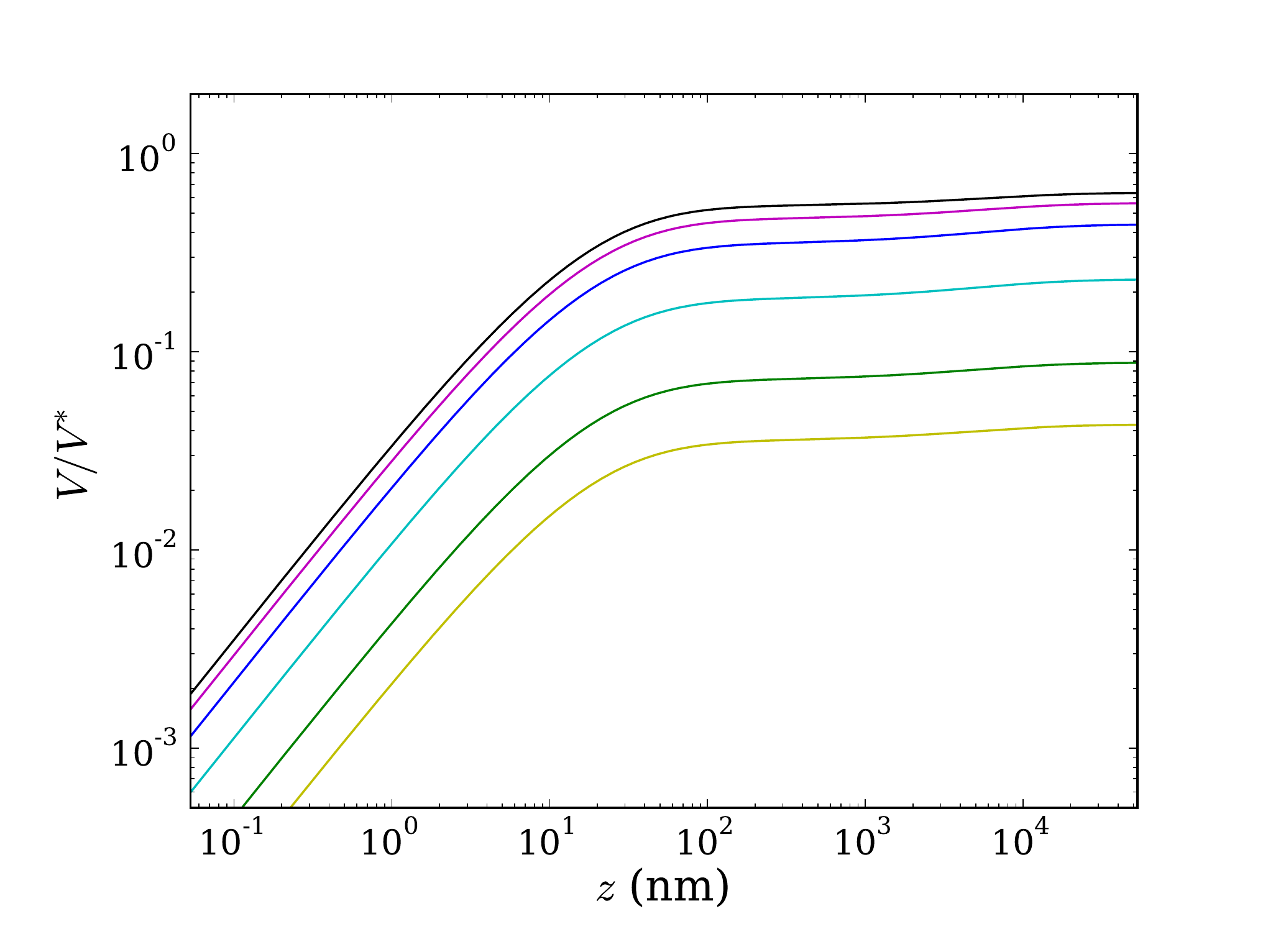}
\caption{\label{vcp-diam}(Color online) CP potential $V$
for antihydrogen on a powder of nano-diamonds, divided by the
long-range potential $V^*$ for a perfect mirror. From top to bottom
the porosity is  0\% (bulk diamond, black), 25\% (magenta), 50\%
(blue), 75\% (cyan), 90\% (green), 95\% (yellow).}
\end{figure}

\begin{figure}[ht]
\centering
\includegraphics[width=8cm]{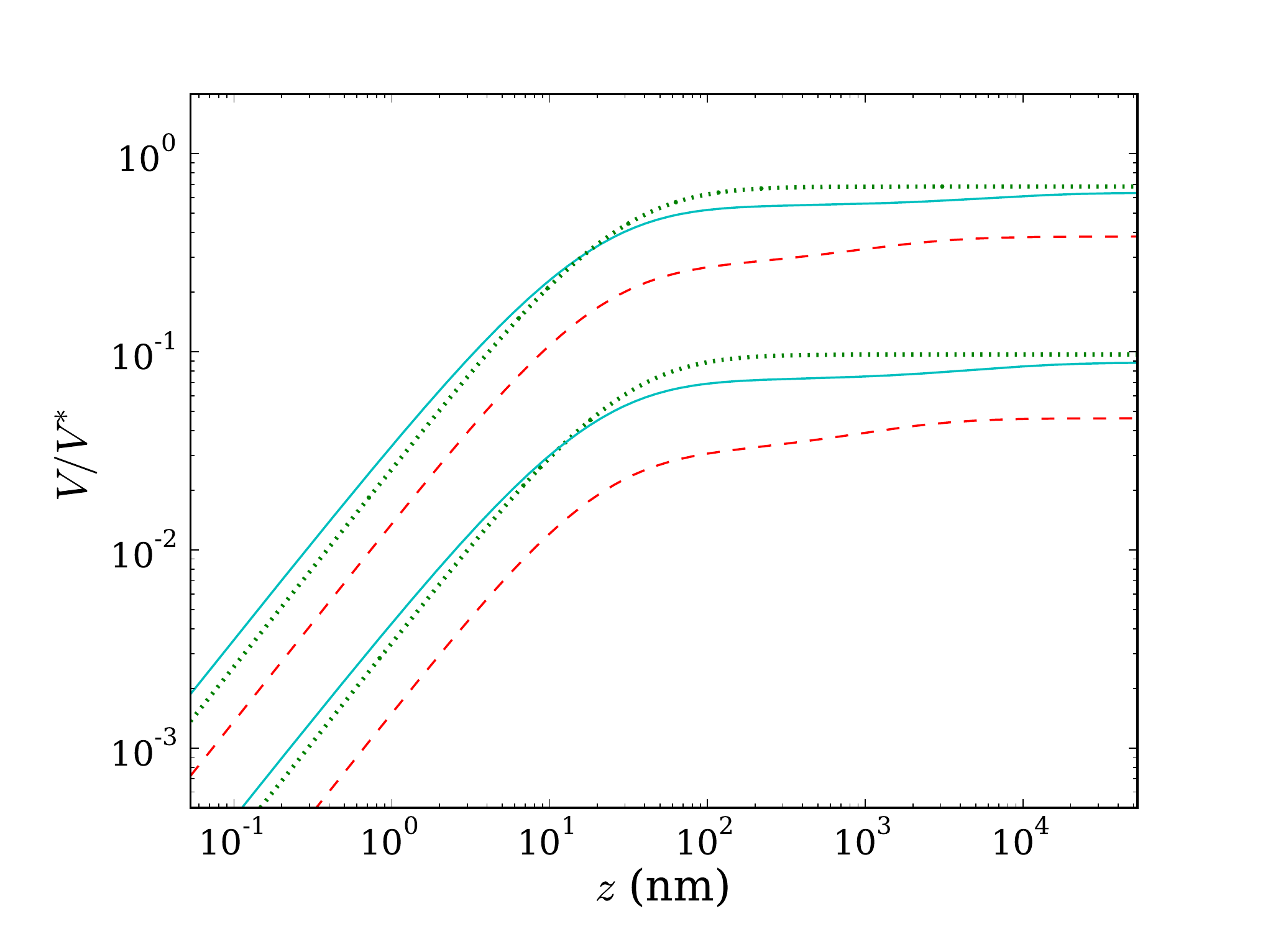}
\caption{\label{vcp-compare}(Color online) CP potential
$V$  for antihydrogen on silica (red dashed line), silicon (green
dotted line) diamond (cyan full line), divided by the long-range
potential $V^*$ for a perfect mirror. The top three curves are for
bulk materials, the bottom three for 90\% porosity. }
%gabriel changed the caption to ressemble the captions of previous figures
\end{figure}

The calculation of the CP potential, presented in
\cite{dufour2012}, is not repeated here. Results are presented with
reference to the long-range retarded potential calculated for a
perfect mirror $V^*(z)= C_4^*/z^4$ with $C_4^* = 2.5 \times
10^{-57}\ \textrm{Jm}^4 = 73.6\ \textrm{a.u.}$~. Figures
\ref{vcp-aero} and \ref{vcp-diam} show the CP
potential created by aerogels and powders of diamond
nanoparticles respectively, for various values of the porosity. 
The potential curves for porous silicon are similar to those of diamond nanoparticles.
Figure \ref{vcp-compare} compares the potentials of these three
materials when the porosity is 0\% and 90\%.

%gabriel changed x-axis units to nanometers instead of atomic units to be more consistent with rest of paper

%\todo[inline]{Figures \ref{vcp-porsi} looks very much like
%\ref{vcp-diam}. Do we drop it for shortening the paper ?}

% \begin{figure}[ht]
% \centering
% \includegraphics[width=8cm]{images/potential-porous-si.eps}
% \caption{\label{vcp-porsi}(Color online) CP potential
% $V$ for antihydrogen on porous silicon, divided by the long-range
% potential $V^*$ for a perfect mirror. From top to bottom the
% porosity is 0\% (bulk silicon, black), 25\% (magenta), 50\% (blue),
% 75\% (cyan), 90\% (green), 95\% (yellow).}
% \end{figure}

%\todo[inline]{Same question for figure \ref{vcp-diam} as the
% essential information is reproduced on figure \ref{vcp-compare}.
% Do we drop it for shortening the paper ?}

%gabriel removed :
% ``The calculation of the Casimir-Polder (CP) potential, presented in
% \cite{dufour2012}, is not repeated here. As explained there, quantum
% reflection occurs in the so-called badlands regions where the
% adiabatic approximation breaks down. 
% Non-adiabatic transitions are effective when the function $Q(z)$, which is related to the
% Schwarzian derivative of the WKB phase, is large (see the references in \cite{dufour2012}).''
% and replaced it with
% ``As explained in \cite{dufour2012}, quantum reflection occurs in the so-called badlands regions where the 
% adiabatic approximation breaks down. 
% Non-adiabatic transitions are effective when the function
% \begin{equation}
% Q(z)  = \frac{\hbar^2}{2}
% \frac{p^{\prime\prime}(z)}{p(z)^3} - \frac{3 \hbar^2}{4}
% \left(\frac{p^{\prime}(z)}{p(z)^2}\right)^2
% \end{equation}
% is large. It is expressed here as a function of the classical momentum $p=\sqrt{2m(E-V(z))}$ and is related to the
% Schwarzian derivative of the WKB phase (see the references in \cite{dufour2012}).''

As explained in \cite{dufour2012}, quantum reflection occurs in the so-called badlands regions where the 
adiabatic approximation breaks down. 
Non-adiabatic transitions are effective when the function
% $
% Q(z)  = \frac{\hbar^2}{2}
% \frac{p^{\prime\prime}(z)}{p(z)^3} - \frac{3 \hbar^2}{4}
% \left(\frac{p^{\prime}(z)}{p(z)^2}\right)^2
% $
$
Q(z)  = \hbar^2 p^{\prime\prime}(z)/ 2 p(z)^3 - 3 \hbar^2
p^{\prime}(z)^2/ 4 p(z)^4
$
is large. It is expressed here as a function of the classical momentum $p=\sqrt{2m(E-V(z))}$ and is related to the
Schwarzian derivative of the WKB phase (see the references in \cite{dufour2012}). 

%gabriel replaced ``peak value'' by ``peak''.
% An alternative would be to say ``peak value and position give an
% indication on the distance where the reflection occurs and on the
% magnitude of the reflection respectively''

\begin{figure}[ht]
\centering
\includegraphics[width=8cm]{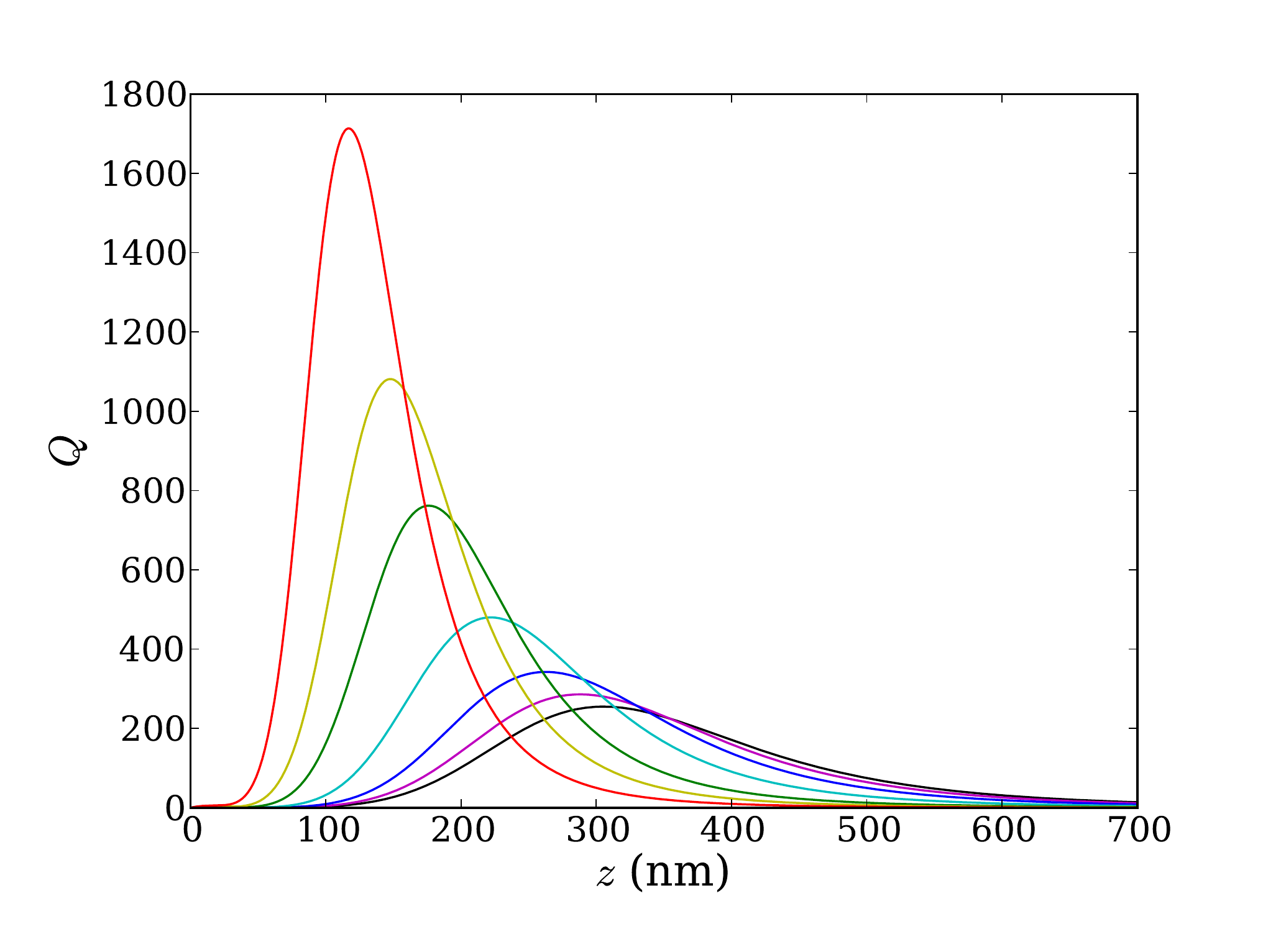}
\caption{(Color online) \label{Qvsz-aerogel} Badlands function $Q$
for antihydrogen approaching silica aerogels with velocity $v=$ 10
mm.s$^{-1}$. From bottom right to top left the porosity is  0\%
(bulk silica, black), 25\% (magenta), 50\% (blue), 75\% (cyan), 90\%
(green), 95\% (yellow), 98\% (red).}
\end{figure}

The peak of this function gives an
indication on the distance where the reflection occurs and on the
magnitude of the reflection. The peak is located closer to the
surface for a weaker potential, thus corresponding to a steeper
variation and a larger $Q$. This is illustrated in figure
\ref{Qvsz-aerogel} for silica aerogels. When porosity is increased,
the peak of the badlands function $Q$ gets closer to the surface
while its magnitude grows.

We note that reflection occurs at a distance $\gtrsim 100\
\textrm{nm}$ if the velocity of atoms is kept below 10
$\textrm{mm.s}^{-1}$, even for the poorest reflective material
considered here (silica aerogel with 98\% porosity). We therefore
except that our simplified model gives qualitatively correct results
as long as the condition on velocity is obeyed. This is the case for
the regime of large quantum reflections obtained at low energies.
%gabriel replaced ``case'' by ``regime'' 

\begin{figure}[ht]
\centering
\includegraphics[width=8cm]{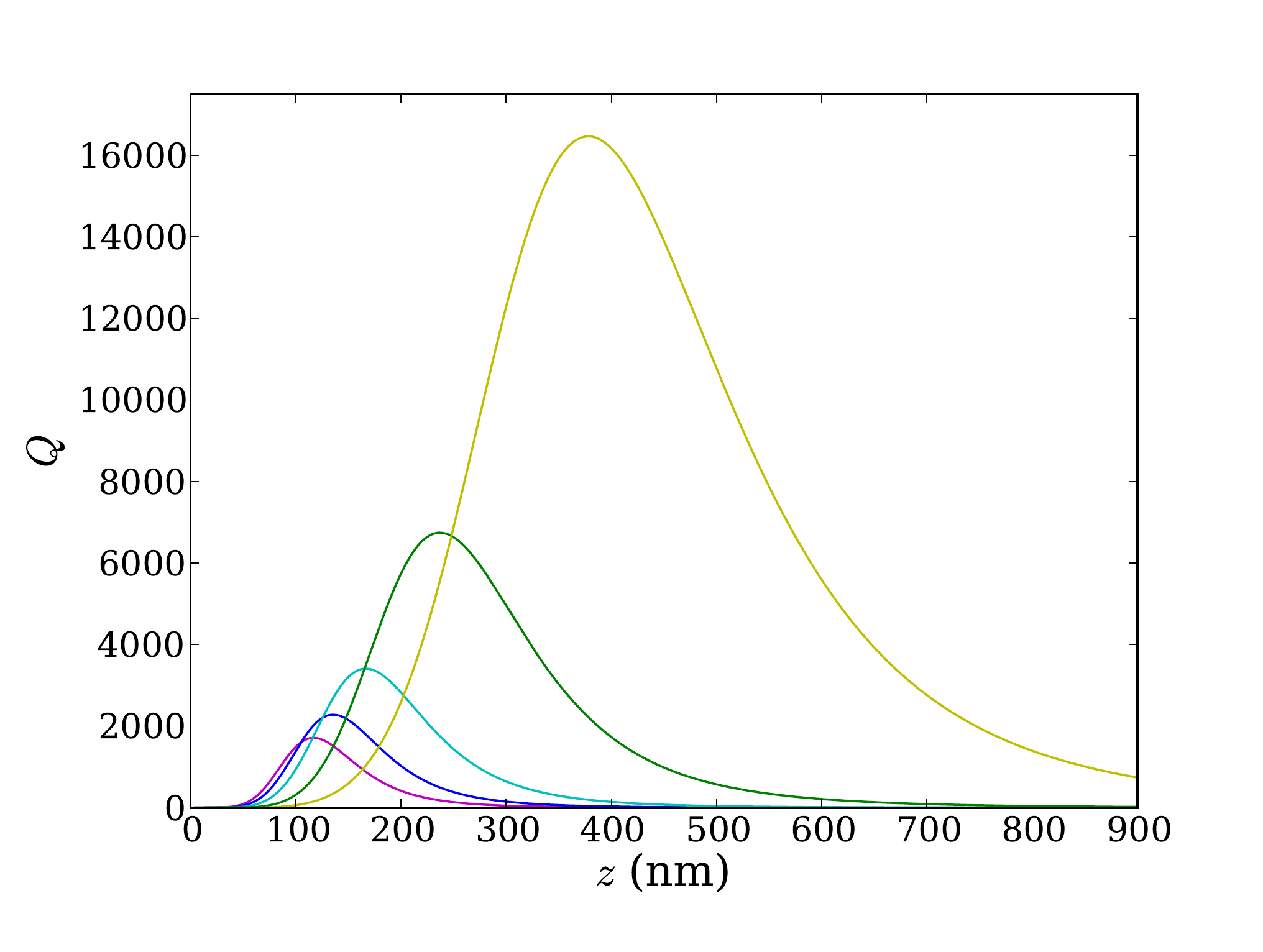}
\caption{(Color online) \label{Qvsz-aerogel-veloc} Badlands function
$Q$ for antihydrogen approaching 98\% porosity silica aerogel. From
bottom left to top right the atom's velocity is  10 mm.s$^{-1}$
(magenta), 7.5 mm.s$^{-1}$ (blue), 5.0 mm.s$^{-1}$ (cyan), 2.5
mm.s$^{-1}$ (green), 1.0 mm.s$^{-1}$ (yellow).}
\end{figure}

\section{Scattering length}

We use the method described in \cite{dufour2012} to calculate the
reflection probability of antihydrogen atoms on the medium. In
particular we enforce a full absorption boundary condition on the
surface to account for the annihilation of those antihydrogen atoms
which come to contact with matter. We focus the discussion on the
low-energy behavior of atoms which corresponds to large quantum
reflection.

\begin{figure}[ht]
\centering
\includegraphics[width=8cm]{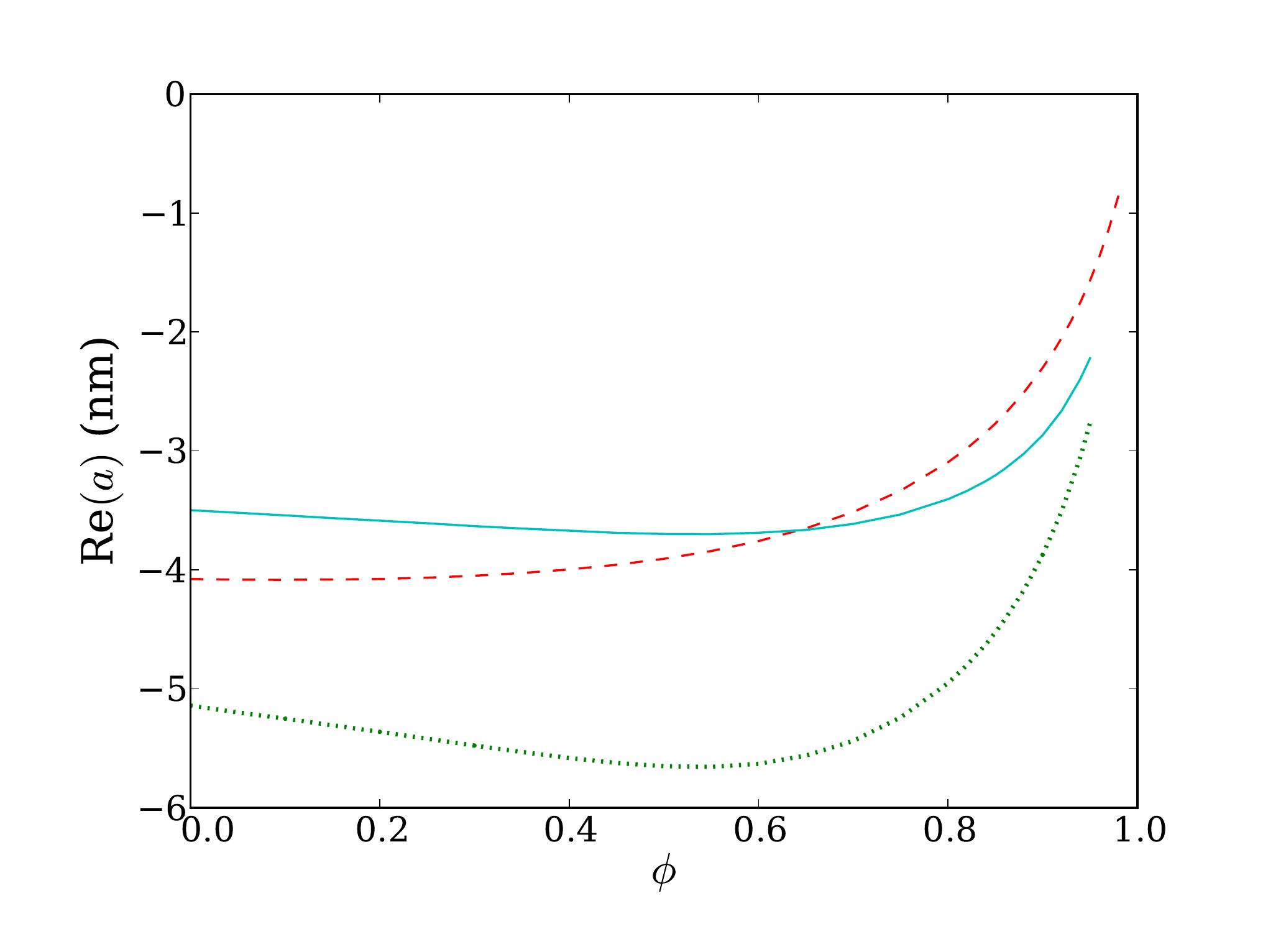}
\caption{\label{rea}(Color online) Real part of the scattering
length $a$ as a function of porosity $\phi$ for antihydrogen on a silica aerogel (red dashed line),
porous silicon (green dotted line) and a powder of nanodiamonds
(cyan full line).}
\end{figure}

\begin{figure}[ht]
\centering
\includegraphics[width=8cm]{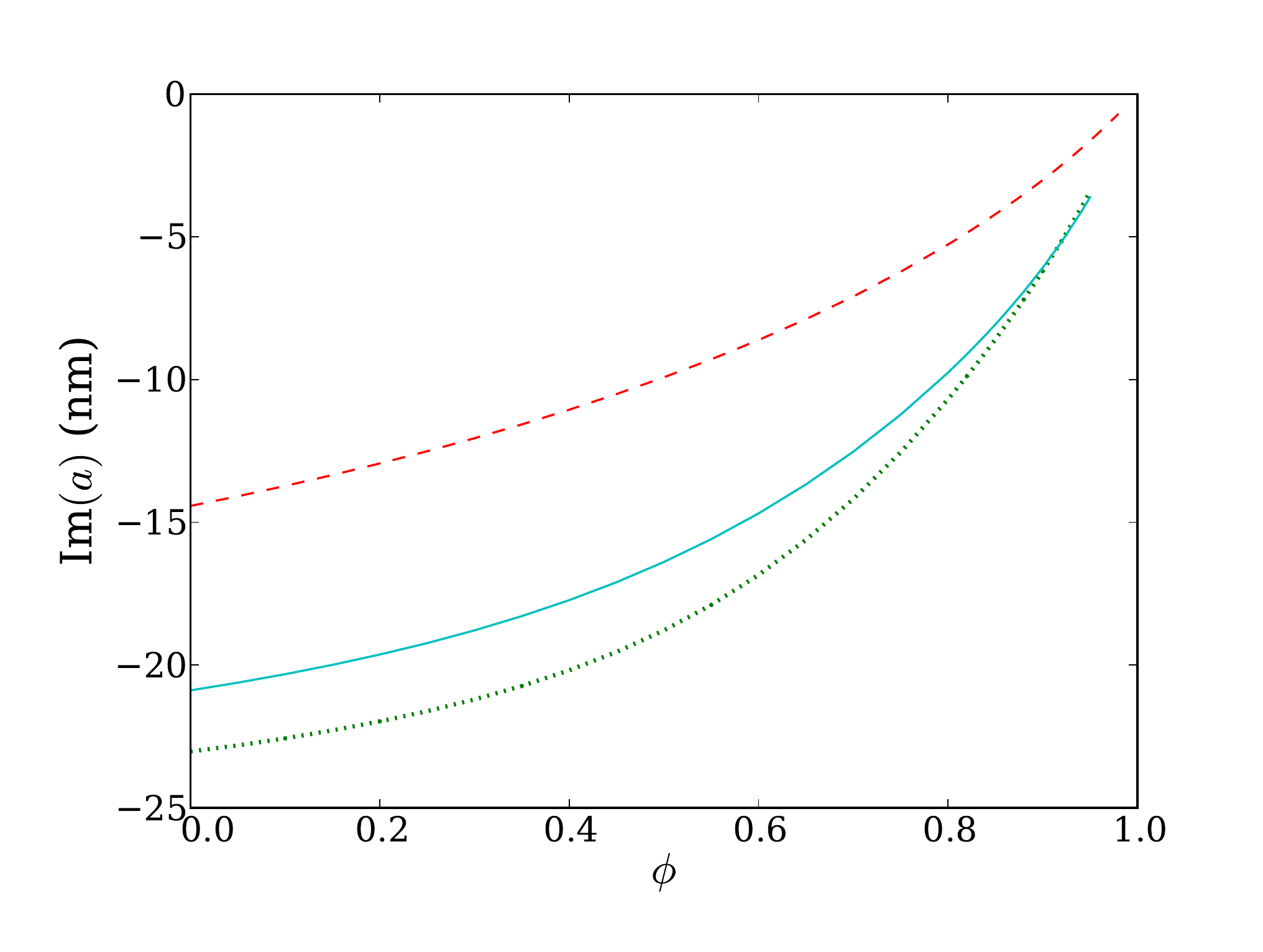}
\caption{\label{ima}(Color online) Imaginary part of the scattering
length $a$ as a function of porosity $\phi$ for antihydrogen on a silica aerogel (red dashed line),
porous silicon (green dotted line) and a powder of nanodiamonds
(cyan full line).}
\end{figure}

The reflection amplitude $r$ can be written in terms of a
complex scattering length $a$, which is defined as in \cite{dufour2012}.  For ultra-cold antihydrogen atoms, that is for
velocities below 10 mm.s$^{-1}$, $a$ is independent of the energy. Its real and imaginary parts are shown in
figures \ref{rea} and \ref{ima} respectively, for varying
porosities.

% The reflection amplitude $r$ is a function of the wavevector
% $k=\sqrt{2mE}/\hbar$ which can be written in terms of a
% complex-valued function $a(k)$ having the dimension of a length
% $r(k) = -\exp\left(-2 ik a(k)\right)$. The real part of $a(k)$
% determines the phase at reflection while its imaginary part
% determines the reflection probability $|r|^2 = \exp\left(4 k
% \mathrm{Im}(a(k) \right)$. $a(k)$ goes to a finite value $a(0)$ when
% $k\to 0$, which is known as the scattering length simply denoted $a$
% in the following. For ultra-cold antihydrogen atoms, that is for
% velocities below 10 mm.s$^{-1}$, $a(k)$ is practically equal to the
% scattering length $a$. Its real and imaginary parts are shown on
% figures \ref{rea} and \ref{ima} respectively, for varying
% porosities.

\begin{figure}[ht]
\centering
\includegraphics[width=8cm]{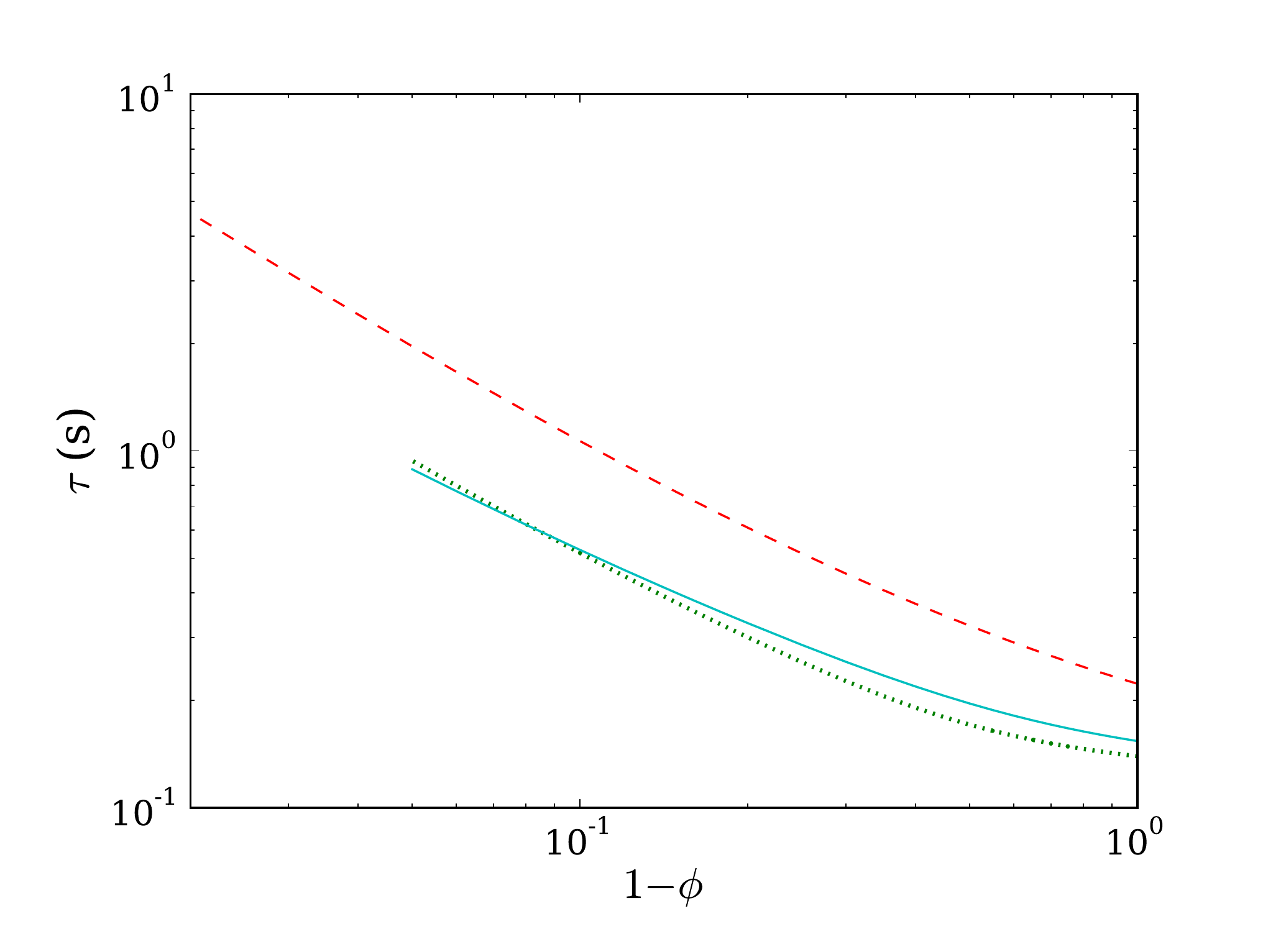}
\caption{\label{lifetime}(Color online) Lifetime $\tau$ for
antihydrogen in the first gravitational state above a silica aerogel
(red dashed line), porous silicon (green dotted line) and a powder
of nanodiamonds (cyan full line), as a function of the solid
fraction $1-\phi$ (log-log scale).}
\end{figure}

When considering quantum gravitational traps for $\Hb$ bounded below
by the quantum reflection from the CP potential and above by gravity
\cite{voronin2011,voronin2012}, one obtains the
lifetime for the quantum bouncer in the first gravitational quantum
state:
\begin{equation}
\tau=\frac{\hbar}{2 m g \left|\mathrm{Im}\,a(0)\right|}~.
\end{equation}
Figure \ref{lifetime} shows this lifetime for silica aerogel, porous
silicon and powder of diamond nanoparticles as the porosity is
varied. The $y-$ axis starts at 0.1~s which is the lifetime
calculated for a perfect mirror \cite{voronin2011}. Reflection is
dramatically enhanced, especially for silica aerogels, for which
extremely high porosities can be reached. For porous silicon and
diamond nanoparticles, the reflection is enhanced by a factor $\sim$
6 between 0\% and 95\% porosity whereas for silica aerogel the
enhancement factor is more than 20 between 0\% and 98\% porosity,
with a lifetime reaching 4.6~s for the latter.

\section{Conclusion}

Using a simple effective medium model, we have shown a dramatic
increase of the quantum reflection probability of antihydrogen atoms
from nanoporous media. We have given theoretical predictions for
reflection on silica aerogels, porous silicon and powders of diamond
nanoparticles over a wide range of porosities. These results open
exciting perspectives for trapping antihydrogen atoms above material
surfaces and investigating its gravitational properties, although more work
is needed to quantify the effects of inhomogeneities in the
nanoporous medium.
%gabriel joined 2 sentences with ``although''

%merlin.mbs 2010-03-15 4.21a (PWD, AO, DPC)
%Control: key (0)
%Control: author (8) initials jnrlst
%Control: editor formatted (1) identically to author
%Control: production of article title (-1) disabled
%Control: page (0) single
%Control: year (1) truncated
%Control: production of eprint (0) enabled
%

\end{document}